\documentclass{ptephy_v1}

\preprintnumber{KUNS-2736}

\usepackage{bm,amssymb, color}
\usepackage{amsmath, amsthm, mathrsfs}
\usepackage{braket}
\usepackage{cases}
\usepackage{txfonts}

\begin{document}

\title{
Ab-initio description 
of excited states of a one-dimensional nuclear
matter with the Hohenberg--Kohn-theorem-inspired
functional-renormalization-group method
}

\author{Takeru Yokota$^{*}$}
\author{Kenichi Yoshida}
\affil{Department of Physics, Faculty of Science, 
Kyoto University, Kyoto 606-8502, Japan
\email{tyokota@ruby.scphys.kyoto-u.ac.jp}}
\author{Teiji Kunihiro}
\affil[2]{Yukawa Institute for Theoretical Physics, 
Kyoto University, Kyoto 606-8502, Japan}



\begin{abstract}
We demonstrate for the first time that a 
functional-renormalization-group aided density-functional theory (FRG-DFT)
describes well the characteristic features of 
the excited states as well as the ground state
of an interacting many-body system
with infinite number of particles
in a unified manner. 
The FRG-DFT is applied to 
a $(1+1)$-dimensional spinless nuclear matter. 
For the excited states, 
the density--density spectral function is calculated
at the saturation point obtained in the framework of FRG-DFT,
and it is found that 
our result reproduces a notable feature of 
the density--density spectral function 
of the non-linear Tomonaga-Luttinger liquid:
The spectral function has a singularity 
at the edge of its support of the lower-energy side.
These findings suggest that the FRG-DFT is
a promising first-principle scheme to analyze the excited states as well as 
the ground states of quantum many-body systems starting 
from the inter-particle interaction. 
\end{abstract}

\subjectindex{D10, B32, A63}

\maketitle

Density-functional theory (DFT) has greatly contributed to our understanding 
of quantum many-body systems in various fields 
including quantum chemistry and atomic, molecular, condensed-matter, and nuclear physics; 
see Refs.~\cite{coh12, lau13, mar17, jon15, nak16,dru09} for some recent reviews. 
The DFT is founded by the Hohenberg-Kohn (HK) theorem~\cite{hoh64}. 
The theorem states that the total energy of the system is a functional 
of the particle density 
which is a function of single variable $\vec{x}$ and that the variational principle with respect to the density 
gives the ground-state density and energy exactly. 
The HK theorem is, however, just an existence theorem,
but the DFT or the HK theorem cannot tell us about the energy-density functional (EDF) 
that we need to minimize. 
The EDFs employed usually in the practical calculations
are thus constructed phenomenologically, and 
improvement of the EDFs lies at the center in the studies based on DFT. 
Therefore, it is highly desirable to develop a systematic method 
to derive the EDF from the underlying microscopic Hamiltonian.

Successful application to the ground state of interacting systems 
in conjunction with the Kohn-Sham theory~\cite{koh65}  
has stimulated attempts to describing 
excited states and dynamics in a framework of time-dependent DFT (TDDFT)~\cite{run84, lau13, nak16}. 
Presently, the linear-response TDDFT has been successfully applied to the small-amplitude 
collective modes of excitation, 
and the real-time TDDFT has been developed to describe even the non-linear dynamics 
as an initial-value problem. 
The TDDFT, in principle, can describe the many-body dynamics exactly. 
It is, however, an open problem to develop a practical method to extract the information 
of excited states possessing the large-amplitude collective character.

In view of quantum field theory, 
the two-particle point-irreducible (2PPI) effective action formalism~\cite{ver92}
gives the HK theorem naturally 
and further the foundation of TDDFT is given in a unified way~\cite{fuk94,val97}. 
Here, the starting point is a generating functional with a source coupled to 
the local composite density operator $\hat{\psi}^\dagger(\vec{x}) \hat{\psi}(\vec{x})$. 
And then a functional Legendre transformation with respect to the source leads to 
an effective action of the density, which gives the quantum equation of motion. 
Therefore, the 2PPI effective action is considered to be a generalization of the EDF.

Let us call here the exact or functional renormalization group (FRG) method~\cite{wet93}, 
which is established as a practical way to treat the effective action non-perturbatively 
and has been successfully applied to 
quantum many-body problems~\cite{ber02, paw07, gie12, met12}. 
The FRG is based on a one-parameter flow equation for 
the parameter-dependent effective average action, 
which gives the effective action of a fully-interacting system eventually 
by taking the quantum fluctuation and correlation gradually starting from a bare system.
The  2PPI effective action formalism combined with the FRG thus gives possibly 
a systematic construction of the EDF based on a microscopic Hamiltonian \cite{pol02, sch04, pol05,bra12, kem13, kem17,kem17b,ram17}; we call such an approach the functional-renormalization-group aided density-functional theory (FRG-DFT).
This approach can be a promising scheme 
for solving the fundamental problems in DFT 
and providing further insights in understanding the many-body systems. 

The FRG-DFT method has been applied to a zero-dimensional model of anharmonic vibrator~\cite{kem13, lia18}, 
and a one-dimensional quantum anharmonic vibrator~\cite{kem13} 
to show the feasibility and effectiveness.  
The estimation of uncertainty due to the truncation was given and an effective way to treat the higher-order 
correction was proposed~\cite{lia18}. 
The FRG-DFT method has also
been applied to a (1+1)-dimensional many-body model 
simulating one-dimensional nucleons
with a fixed particle-number formalism~\cite{kem17}, 
although the bound-state energy was underestimated by 
approximately 30\% in comparison with 
the exact solution in two-particle system
and over 17\% in comparison with 
the results obtained using the Monte Carlo method \cite{ale89} when the number of particles is no more than eight.
In addition, the study of (1+1)-dimensional systems composed of finite number of spin-1/2 fermions interacting via a contact interaction has been reported \cite{kem17b,ram17}.

Since the 2PPI effective action is an effective action which is by definition capable of describing 
 time-dependent phenomena as well as static phenomena equally,
the FRG-DFT method should be in principle applicable 
to not only the ground state but also the excited sates
though there have been no attempts to demonstrate it 
as far as we are aware of.
It should be noted here that the FRG has been applied 
to obtain the spectral functions in the $O(N)$ model~\cite{kam14} 
and the quark-meson model~\cite{tri14a, tri14b, yok16, yok17}. 
Here, the analytic continuation is taken at each order before evaluation of the flow equations.
This technique is thus much easier numerically than the standard method 
such as the maximum entropy method or the Pad\'e approximation.



In this Letter, we demonstrate for the first time 
that our FRG-DFT works well
for describing the excited states as well as 
the ground states of continuum matter.
We are going to consider a (1+1)-dimensional
spinless nuclear matter~\cite{ale89} with infinite number of particles.
After summarizing our result for the ground state energy \cite{yok18}
where the resultant equation of state was found to give the saturation energy
compatible with that obtained using the Monte Carlo method \cite{ale89},
we show the numerical result for the density--density spectral function.
We find that our density--density
spectral function reproduces the existence
of the peak at the edge of its support in the lower-energy
side, which is known as a notable feature of
the non-linear Tomonaga--Luttinger liquid.
Our result suggests that the FRG-DFT is
a powerful way to analyze excited states 
as well as ground states of quantum many-body systems.

For self-containedness,we first recapitulate some part of our FRG-DFT formalism 
to analyze the ground state properties of infinite matters,
which has been developed by the present authors in Ref.~\cite{yok18}.

A microscopic input in our study is the inter-particle interaction. 
The interaction we adopted consists of the short-range repulsive core 
and the long-range attractive force of `nucleons' both of which are given by a Gaussian :
	$U(r)
	=
	({g}/\sqrt{\pi})(\sigma_1^{-1}{\rm e}^{-r^2/\sigma_1^2}
	-
	\sigma_2^{-1}{\rm e}^{-r^2/\sigma_2^2})$,
where $\sigma_1 >0$, $\sigma_2 >0$ and $g >0$.
As given in Ref.~\cite{ale89}, 
we chose $g=12$, $\sigma_1=0.2$ and $\sigma_2=0.8$
in units such that the mass of a nucleon is 1.
These values were determined under the assumptions
that the relevant dimensionless values in one dimension
are comparable with the empirical ones in three dimensions.

Although we are primarily interested in the system with 
zero-temperature in the present work,
we found that the use of the finite-temperature 
imaginary-time formalism is most convenient.
Then the action of the one-dimensional 
interacting spinless fermions reads
\begin{align}
	&S[\psi^*,\psi]
	=
	\int_{\chi}
	\psi^* (\chi)\left(
	\partial_\tau-\frac{1}{2}\partial^2_x
	\right)\psi(\chi)
	+
	\frac{1}{2}\int_{\chi,\chi'}
	\psi^* (\chi)\psi^* (\chi')
	U_{\rm 2b}(\chi,\chi')
	\psi(\chi')\psi(\chi),
	\label{eq:action}
\end{align}
where we have introduced the shorthands
 $\chi=(\tau,x)$, $\int_{\chi}=\int_{-\beta/2}^{\beta/2}d\tau\int dx$ 
 and $U_{\rm 2b}(\chi,\chi'):=U_{\rm 2b}(\chi-\chi'):=\delta(\tau-\tau')U(x-x')$.

Since we are going to employ the techniques developed in the FRG method, 
we regulated the interaction between
fermions by multiplying $U_{\rm 2b}(\chi,\chi')$
by a regulator function $\mathcal{R}_{\lambda}(\tau,x,\tau',x')$ as introduced in Refs.~\cite{pol02,sch04,kem17}, 
and the resulting regulated action is given as
\begin{align}
	&
	S_{\lambda}[\psi^*,\psi]
	=
	\int_{\chi}
	\psi^* (\chi)\left(
	\partial_\tau-\frac{1}{2}\partial^2_x
	\right)\psi(\chi)
	+
	\frac{1}{2}\int_{\chi,\chi'}
	\psi^* (\chi)\psi^* (\chi')
	\mathcal{R}_{\lambda}(\chi,\chi')
	U_{\rm 2b}(\chi,\chi')
	\psi(\chi')\psi(\chi).
	\label{eq:actionreg}
\end{align}
Here, $\mathcal{R}_{\lambda}(\chi,\chi')$ was chosen so as to satisfy
the following conditions:
$\lim_{\lambda\rightarrow 0}
\mathcal{R}_{\lambda}(\chi,\chi')=0$ and 
$\lim_{\lambda\rightarrow 1}
\mathcal{R}_{\lambda}(\chi,\chi')=1$.
Under these conditions,
$S_{\lambda}$ becomes the action of free particles at $\lambda=0$
and that of interacting particles at $\lambda=1$, namely Eq.~\eqref{eq:action}.
Specifically, we chose 
$\mathcal{R}_\lambda (\chi_1,\chi_2)=\lambda$ 
for simplicity~\cite{pol02,sch04}.
We then define the $\lambda$-dependent generating functional 
for density-density correlation functions as
$Z_{\lambda}[J]
=\int \mathcal{D}\psi^* \mathcal{D}\psi 
\exp
(
-S_{\lambda}[\psi^*,\psi]
+\int_{\chi}J(\chi)\hat{\rho}(\chi))$
with $\hat{\rho}(\chi)=\psi^*(\chi)\psi(\chi)$ being the composite 
local number density operator.
The generating functional for the connected density correlation functions
$G_{\lambda}^{(n)}(\chi_1,\cdots,\chi_n)$ is given as
$W_{\lambda}[J]=\ln Z_{\lambda}[J]$,
i.e. $G_{\lambda}^{(n)}(\chi_1,\cdots,\chi_n)
=\delta^{n} W_\lambda [J]/\delta J(\chi_1)\cdots\delta J(\chi_n)|_{J=0}$.

Then the effective action $\Gamma_\lambda[\rho]$ of the local density $\rho(\chi)$
is obtained by the Legendre transformation of $W_{\lambda}[J]$:
$\Gamma_{\lambda}[\rho]=\sup_{J}
( -W_{\lambda}[J]+\int_\chi J(\chi)\rho(\chi))$.
An important feature of $\Gamma_\lambda[\rho]$
is that it can be related to the energy density
functional $E_\lambda[\rho]$ 
as $E_{\lambda}[\rho]=\lim_{\beta\rightarrow \infty}\Gamma_{\lambda}[\rho]/\beta$ \cite{fuk94},
i.e. the ground state density $\rho_{\rm gs,\lambda}(\chi)$ and energy $E_{\rm gs,\lambda}$
are obtained variationally from $\Gamma_{\lambda}[\rho]$.
When considering the variational problem
under the constraint that the particle number is 
set to some value,
we should minimize 
$I_\lambda[\rho]:=\Gamma_\lambda[\rho]-\mu_\lambda\int_\chi\rho(\chi)$
with respect to $\rho(\chi)$.
Here, we have introduced a $\lambda$-dependent
chemical potential $\mu_\lambda$
to control the change of 
the particle number during the flow
caused by switching on of the interaction \cite{yok18}.
In this case, the ground state density 
$\rho_{\rm gs,\lambda}(\chi)$
satisfies the following stationary condition:
$(\delta\Gamma_\lambda/\delta \rho(\chi))[\rho_{\rm gs,\lambda}]=\mu_\lambda$.
Here, we should mention that 
the chemical potential 
depending on the RG parameter was studied~\cite{hon01,vil17}
in the framework of
the functional RG \`a la Wetterich~\cite{wet93}
and the change of the chemical potential
by the presence of interaction was discussed 
in the context of DFT \cite{dru09}.

The renormalization group flow 
equation of $\Gamma_\lambda [\rho]$
reads \cite{sch04,kem17}
\begin{align}
	\partial_\lambda \Gamma_\lambda [\rho]
	=
	\frac{1}{2}\int_{\chi_1,\chi_2}
	U_{\rm 2b}(\chi_1,\chi_2)
	\left(
	\rho (\chi_1)
	\rho(\chi_2)
	+
	\left(
	\frac{\delta^2 \Gamma_\lambda [\rho]}{\delta \rho \delta \rho}\right)^{-1}(\chi_1,\chi_2)
	-\rho(\chi_2)\delta(x_2-x_1)
	\right).
	\label{eq:masterfloweq}
\end{align}
One can calculate the effective action 
$\Gamma_{\lambda=1}[\rho]$, 
whose classical action is given in Eq.~\eqref{eq:action},
by solving Eq.~\eqref{eq:masterfloweq}
with the initial condition 
$\Gamma_{\lambda=0}[\rho]$, which is
the effective action of the non-interacting system.
The functional differential equation
\eqref{eq:masterfloweq} can be 
converted
to an infinite series of differential equations
by the expansion around $\rho=\rho_{\rm gs,\lambda}$.
In particular, from Eq.~\eqref{eq:masterfloweq}
and its second derivative around 
$\rho=\rho_{\rm gs,\lambda}$, and the stationary condition,
the flow equations of
the energy $E_{\rm gs,\lambda}$,  
the density $\rho_{\rm gs, \lambda}$ at the ground state,
and the two-point correlation function
$G_{\lambda}^{(2)}(\chi,\chi')$
are obtained as follows:
\begin{align}
	\partial_\lambda E_{\rm gs,\lambda}
	=&
	\lim_{\beta\to \infty}
	\frac{1}{\beta}\left[\int_\chi
	\mu_\lambda
	\partial_\lambda\rho_{\rm gs,\lambda}(\chi)
	\right.
	\notag
	\\
	&+
	\left.\frac{1}{2}
	\int_{\chi,\chi'}
	U_{\rm 2b}(\chi,\chi')
	\left(
	\rho_{\rm gs, \lambda}(\chi)
	\rho_{\rm gs, \lambda}(\chi')
	+
	G_{\lambda}^{(2)}(\chi,\chi')
	-\rho_{\rm gs,\lambda}(\chi')\delta(x'-x)
	\right)\right],
	\label{eq:fflo}
	\\
	\partial_\lambda \rho_{\rm gs,\lambda}(\chi)
	=&
	-\frac{1}{2}\int_{\chi_1,\chi_2}
	U_{\rm 2b}(\chi_1,\chi_2)
	G_{\lambda}^{(3)}(\chi_2,\chi_1,\chi)
	\notag
	\\
	&+
	\int_{\chi_1}G_{\lambda}^{(2)}(\chi,\chi_1)
	\left(
	\partial_\lambda \mu_\lambda
	-
	\int_{\chi_2}
	U_{\rm 2b}(\chi_1,\chi_2)
	\rho_{\rm gs,\lambda}(\chi_2)
	+
	\frac{1}{2}U(0)
	\right),
	\label{eq:rflo}
	\\
	\partial_\lambda G_{\lambda}^{(2)}(\chi,\chi')
	=&
	-\int_{\chi_1,\chi_2}
	U_{\rm 2b}(\chi_1,\chi_2)
	\left(
	G_{\lambda}^{(2)}(\chi,\chi_1)
	G_{\lambda}^{(2)}(\chi_2,\chi')
	+\frac{1}{2}
	G_{\lambda}^{(4)}(\chi_2,\chi_1,\chi,\chi')
	\right)
	\notag
	\\
	&+
	\int_{\chi_1}G_{\lambda}^{(3)}(\chi,\chi',\chi_1)
	\left(
	\partial_\lambda \mu_\lambda
	-
	\int_{\chi_2}
	U_{\rm 2b}(\chi_1,\chi_2)
	\rho_{\rm gs,\lambda}(\chi_2)
	+
	\frac{1}{2}U(0)
	\right).
	\label{eq:g2fl}
\end{align}

In this Letter, we assume that the ground state of 
the system is homogeneous for any $\lambda$.
In this case, we can set $\rho_{\rm gs,\lambda}$
to a constant value during the flow,
i.e. $\partial_\lambda\rho_{\rm gs,\lambda}=0$,
by choosing $\mu_\lambda$ so as to satisfy
$
\partial_\lambda \mu_\lambda
=
\tilde{U}(0)
\rho_{\rm gs,\lambda}
-U(0)/2
+\int_{P}
\tilde{U}(p)\tilde{G}_{\lambda}^{(3)}(P,-P)/
(2\tilde{G}_{\lambda}^{(2)}(0))
$.
Here, for convenience 
we have introduced the momentum representations
$\tilde{U}(p):=\int_{x}U(x)e^{-ipx}$
and
$(2\pi)^2\delta(P_1+\cdots+P_n)
\tilde{G}_\lambda^{(n)}(P_1,\cdots,P_{n-1})
:=
\int_{\chi_1,\cdots,\chi_n}
e^{-i(P_1\cdot\chi_1
+\cdots+P_n\cdot\chi_n)}
G_\lambda^{(n)}(\chi_1,\cdots,\chi_{n})$,
where $P_{i}:=(\omega_{i},p_i)$ is a vector of
a Matsubara frequency and a momentum,
and the short hand
$\int_{P}:=\int dp d\omega/(2\pi)^2$. 
We note that $\tilde{G}_\lambda^{(2)}(0)$ is interpreted as the $p$ limit of 
$\tilde{G}_\lambda^{(2)}(P)$, i.e. 
$\lim_{p\to 0}\tilde{G}_\lambda^{(2)}(0,p)$, 
in our case and thus is regarded as the static 
particle-density susceptibility \cite{for75,kun91,fuj04,yok18}, 
which is usually nonzero.
Under the choice of $\mu_\lambda$, Eqs.~\eqref{eq:fflo} and \eqref{eq:g2fl}
are reduced to the following equations, respectively \cite{yok18}:
\begin{align}
	\partial_\lambda \overline{E}_{\rm gs,\lambda}
	=&
	\frac{\rho_{\rm gs,0}}{2} \tilde{U}(0)
	+
	\frac{1}{2\rho_{\rm gs,0}}
	\int_{p}
	\tilde{U}(p)
	\left(
	\int_{\omega_{\rm}}
	\tilde{G}^{(2)}_{\lambda}(P)
	-
	\rho_{\rm gs,0}
	\right),
	\label{eq:finalenergyflow}
	\\
	\partial_\lambda \tilde{G}_{\lambda}^{(2)}(P)
	=&
	-\tilde{U}(p) \tilde{G}_{\lambda}^{(2)}(P)^2
	-
	\frac{1}{2}
	\int_{P'}
	\tilde{U}(p')
	\left[
	\tilde{G}^{(4)}_{\lambda}(P',-P',P)
	-
	\frac{
	\tilde{G}_{\lambda}^{(3)}(P',-P')
	\tilde{G}_{\lambda}^{(3)}(P,-P)
	}{\tilde{G}_{\lambda}^{(2)}(0)}
	\right]
	\label{eq:finalg2flow},
\end{align}
where we have introduced the energy per particle 
$\overline{E}_{\rm gs,\lambda}=E_{\rm gs,\lambda}/\int dx \rho_{\rm gs,0}$.


Equation~\eqref{eq:finalg2flow} contains 
$\tilde{G}^{(3,4)}_{\lambda}$,
the flow equations for 
which are derived from Eq.~\eqref{eq:masterfloweq} 
in terms of $\tilde{G}^{(n\geq 5)}_{\lambda}$ and so on 
because the flow equation for 
$\tilde{G}^{(n)}_{\lambda}$ depends on
$\tilde{G}^{(2)}_{\lambda},\cdots,
\tilde{G}^{(n+2)}_{\lambda}$.
Thus it is obvious that a truncation scheme is necessary 
for solving the flow equations in a practical calculation.
In the present calculation, we did not consider the flows of $\tilde{G}^{(3,4)}_{\lambda}$.
However, the simple replacement of $\tilde{G}^{(3,4)}_{\lambda}$
by $\tilde{G}^{(3,4)}_{0}$ in Eq.~\eqref{eq:finalg2flow}
causes the breaking of the Pauli exclusion principle.
To avoid this artifact, 
we used the following approximation as introduced in Ref.~\cite{kem17}:
\begin{align}
	&
	\int_{P'}
	\tilde{U}(p')
	\left[
	\tilde{G}^{(4)}_{\lambda}(P',-P',P)
	-
	\tilde{G}_{\lambda}^{(3)}(P',-P')
	\tilde{G}_{\lambda}^{(3)}(P,-P)\tilde{G}_{\lambda}^{(2)}(0)^{-1}
	\right]
	\notag
	\\
	\approx
	&
	f_{\cal P}(\lambda)
	\int_{P'}
	\tilde{U}(p')
	\left[
	\tilde{G}^{(4)}_{0}(P',-P',P)
	-
	\tilde{G}_{0}^{(3)}(P',-P')
	\tilde{G}_{0}^{(3)}(P,-P)\tilde{G}_{0}^{(2)}(0)^{-1}
	\right],
	\label{eq:G4approx}
\end{align}
with $f_{\cal P}(\lambda)$ being a factor
to preserve the effect of Pauli blocking.
According to the Pauli blocking, we have
$\partial_{\lambda}G_\lambda^{(2)}(\chi,\chi)=0$.
Therefore $f_{\cal P}(\lambda)$
is determined using Eq.~\eqref{eq:finalg2flow}:
$
f_{\cal P}(\lambda)
=
-2\int_{P}\tilde{U}(p) \tilde{G}_{\lambda}^{(2)}(P)^2/
\int_{P',P''}
\tilde{U}(p')
[
\tilde{G}^{(4)}_{0}(P',-P',P'')
-
\tilde{G}_{0}^{(3)}(P',-P')
\tilde{G}_{0}^{(3)}(P'',-P'')\tilde{G}_{0}^{(2)}(0)^{-1}].
$
At $\lambda=0$, we have $f_{\cal P}(0)=1$.

To solve the flow equations
\eqref{eq:finalenergyflow} and
\eqref{eq:finalg2flow},
we need the initial conditions $\overline{E}_{\rm gs, \lambda=0}$,
$\rho_{\rm gs,\lambda=0}$, and $\tilde{G}^{(2,3,4)}_{\lambda=0}$.
We denote $\rho_{\rm gs,0}$ by $n$, which
is always the density of the ground state during the flow,
and in particular at $\lambda=1$,
because $\rho_{\rm gs,\lambda}(\chi)=\rho_{\rm gs,0}$.
Then the Fermi momentum and Fermi energy are defined as $p_{\rm F}=\pi n$
and $E_{\rm F}=p_{\rm F}^2/2$, respectively.
$\overline{E}_{\rm gs, \lambda=0}$
is the ground state energy 
of the one-dimensional free Fermi gas:\,$\overline{E}_{\rm gs,0}=E_{\rm F}/3$.
$G^{(2,3,4)}_{\lambda=0}$ are the correlation functions for free particles:
\begin{align}
	G^{(2)}_{0}(P)
	=&
	-
	\int_{P'}
	G_{\rm F, 0}^{(2)}(P+P')
	G_{\rm F, 0}^{(2)}(P'),
	\label{eq:G20}
	\\
	\tilde{G}^{(3)}_{0}(P_1,P_2)
	=&
	-
	\sum_{\sigma\in S_{2}}
	\int_{P'}
	\tilde{G}_{\rm F, 0}^{(2)}(P')
	\tilde{G}_{\rm F, 0}^{(2)}(P_{\sigma(1)}+P')
	\tilde{G}_{\rm F, 0}^{(2)}(P_{\sigma(1)}+P_{\sigma(2)}+P'),	
	\label{eq:G30}
	\\
	G^{(4)}_{0}(P_1,P_2,P_3)
	=&
	-
	\sum_{\sigma\in S_{3}}
	\int_{P'}
	G_{\rm F, 0}^{(2)}(P')
	G_{\rm F, 0}^{(2)}(P_{\sigma(1)}+P')
	G_{\rm F, 0}^{(2)}(P_{\sigma(1)}+P_{\sigma(2)}+P')
	\notag
	\\
	&\times
	G_{\rm F, 0}^{(2)}(P_{\sigma(1)}+P_{\sigma(2)}+P_{\sigma(3)}+P').
	\label{eq:G40}
\end{align}
Here $S_{2}$ and $S_{3}$ are the symmetric groups of order two and three, respectively,
and $G_{\rm F, 0}^{(2)}(P)$ is the two-point propagator
of free fermions:\,$G_{\rm F, 0}^{(2)}(P)={1}/(-i\omega+\epsilon(p))$,
where $\epsilon(p):=p^2/2-E_{\rm F}$.
Using Eqs.~\eqref{eq:G20}-\eqref{eq:G40},
the expressions of the flow equations \eqref{eq:finalenergyflow} and \eqref{eq:finalg2flow}
under the approximation Eq.~\eqref{eq:G4approx} 
are found to be the same as those obtained from the continuum limit
of the system with finite number of particles in a finite box presented in Ref.~\cite{kem17}.

We need to evaluate the 
momentum integrals such as $\int dp' \tilde{U}(p')/p'$, 
which appear in the second term in the right-hand side 
of Eq.~\eqref{eq:finalg2flow}.
The integrand apparently has a singular point at $p'=0$,
which is actually absent because $\tilde{U}(0)=0$ in the present case.
In order to avoid a division-by-zero operation,
we rewrote the integrand by using the Maclaurin expansion of $\tilde{U}(p')$ 
at small $p'$ to a manifestly regular form
for the numerical calculation.

The results for the equation of state and the saturation energy,
and the comparison with the Monte Carlo simulation \cite{ale89}
were shown in Ref.~\cite{yok18}.
A remark is in order here: In the Monte Carlo simulation,
the saturation energy was mere an extrapolated energy at a given density $n=1.16$,
which is considered to be close \cite{ale89} but not equal to the saturation density; 
moreover the extrapolation was made from the results for 
finite particle systems with a particle number up to 12.
In contrast, our FRG-DFT calculation was made for the system
with infinite number of particles, and the density was varied continuously.
The saturation energy derived from FRG-DFT
is quite close to the result of the Monte Carlo simulation:
We found that the discrepancy between the saturation energy given by FRG-DFT
and that by the Monte Carlo simulation is 2.7\%.
We pointed out that such an accuracy
was acquired with little computational time or resources
in our framework of FRG-DFT.


This successful application to the ground-state properties of a many-body system 
rouses one's interest in extension of FRG-DFT to describing excited states. 
Then we are going to describe our way to calculate the density--density spectral function.
We define the density--density spectral function
$\rho_{\rm d,\lambda}(\omega,p)$ as
$\rho_{\rm d,\lambda}(\omega, p)
=-2\mathrm{Im} \tilde{G}_{\rm R,\lambda}^{(2)}(\omega, p)$,
where $\tilde{G}^{(2)}_{\rm R,\lambda}(\omega, p)$
is the retarded two-point density correlation function,
which is obtained from the analytic continuation
of $\tilde{G}^{(2)}_{\lambda}(\omega_{\rm I}, p)$:
$\tilde{G}^{(2)}_{\rm R,\lambda}(\omega, p)
=
-\tilde{G}^{(2)}_{\rm ana,\lambda}(\omega+i\epsilon, p)$.
Here, $\epsilon$ is a positive infinitesimal and
$\tilde{G}^{(2)}_{\rm ana,\lambda}(z, p)$
is a complex function of $z\in\mathbb{C}$ which 
is regular in the upper-half plane of $z$
and satisfies 
$\tilde{G}^{(2)}_{\rm ana,\lambda}(-i\omega_{\rm I}, p)=
\tilde{G}^{(2)}_{\lambda}(\omega_{\rm I}, p)$
for $\omega_{\rm I}\in \mathbb{R}$.
The analytic continuation to obtain 
$\tilde{G}^{(2)}_{\rm ana,\lambda}(z, p)$
from $\tilde{G}^{(2)}_{\lambda}(\omega_{\rm I}, p)$
is often an obstruction for a numerical analysis.
In our case, however,
the analytic continuation 
can be performed in the level of the flow equations
as in Refs.~\cite{kam14,tri14a,tri14b,yok16,yok17},
which is much easier numerically 
than the standard procedures
such as the maximum entropy method or 
the P\'ade approximation.
Under the approximation Eq.~\eqref{eq:G4approx},
one finds that the second term of Eq.~\eqref{eq:finalg2flow}
is regular in the upper-half plane of $z$
when $P^0$ is simply replaced with $-iz$.
Then $\tilde{G}_{\lambda}^{(2)}(P)|_{P^{0}
\rightarrow -iz}$ 
also keeps regular in the upper-half of $z$ during the flow.
Therefore, we have the flow equation for $\tilde{G}^{(2)}_{\rm R,\lambda}(\omega, p)$:
\begin{align}
	\partial_\lambda \tilde{G}_{\rm R,\lambda}^{(2)}(\omega, p)
	\approx
	&\tilde{U}(p) \tilde{G}_{\rm R,\lambda}^{(2)}(\omega, p)^2
	\notag
	\\
	&+
	\frac{f_{\cal P}(\lambda)}{2}
	\int_{P'}
	\tilde{U}(p')
	\left.
	\left[
	\tilde{G}^{(4)}_{0}
	(P',-P',P)
	-
	\frac{
	\tilde{G}_{0}^{(3)}(P',-P')
	\tilde{G}_{0}^{(3)}(P,-P)
	}{\tilde{G}_{0}^{(2)}(0)}
	\right]
	\right|_{P^{0}\rightarrow -i(\omega+i\epsilon)}
	\label{eq:realg2flow}.
\end{align}
The initial condition of this flow equation $\tilde{G}_{\rm R,0}^{(2)}(\omega, p)$
is given by the replacement of $P^0$
with $-i(\omega+i\epsilon)$ in Eq.~\eqref{eq:G20}.
As discussed below, the contribution from the second term in the right-hand side
of the flow equation \eqref{eq:realg2flow} is important to capture the feature of 
the spectral function in (1+1) dimensions.
If the second term of the right-hand side of Eq.~\eqref{eq:realg2flow}
is neglected, this flow equation can be solved analytically: 
We have
$\tilde{G}_{\rm R,\lambda =1}^{(2)}(\omega, p)
=
(\tilde{G}_{\rm R,0}^{(2)}(\omega, p)^{-1}-U(p))^{-1}$,
which is equivalent to that derived in the random phase approximation (RPA).

\begin{figure}[!t]
	\centering
	\includegraphics[width=0.85\columnwidth]{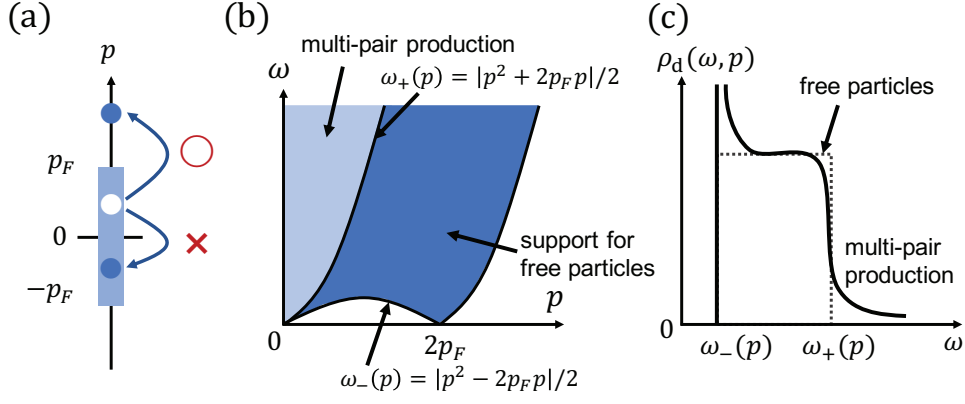}
	\caption
	{(a) Schematic picture of the particle--hole
	excitation from the Fermi sphere. Due to the Pauli blocking, some excitations are forbidden kinematically, which gives the lower bound of
	the energy of the support of 
	the density--density spectral function
	$\omega_{-}(p)$
	in $0\leq p \leq p_{\rm F}$.
	(b) The supports of the spectral functions
	for the interacting case and free case.
	The support in $\omega>\omega_{+}(p)$ is due to the contribution
	from multi-pair productions.
	(c) The expected $\omega$ dependence of $\rho_{\rm d}(\omega,p)$ at a fixed $p$
	in the interacting case (solid line)
	and free case (dotted line).
	\label{fig:concept_rho}}
\end{figure}

Before presenting our result
of the density--density spectral function $\rho_{\rm d}(\omega,p)$,
we briefly mention the expected behavior of the
density--density spectral function of (1+1)-dimensional
interacting fermions.
First, let us consider the free fermion case. 
If the particle--hole excitation
with an energy $\omega$ and a momentum $p$
is kinematically forbidden, 
$\rho_{\rm d}(\omega,p)$ is zero. 
For (1+1)-dimensional 
free fermions,
a particle--hole excitation
with an energy $\omega$ and a momentum $p$
is kinematically allowed
if the following condition 
is satisfied:
$\omega_{-}(p) \leq \omega \leq \omega_{+}(p)$,
where $\omega_{-}(p):=|p^2-2p_{\rm F}p|/2$
and $\omega_{+}(p):=|p^2+2p_{\rm F}p|/2$; 
see Fig.~\ref{fig:concept_rho}(a).
Therefore $\rho_{\rm d}(\omega,p)$ has its support
as shown in Fig.~\ref{fig:concept_rho}(b).
On this support, the strength of $\rho_{\rm d}(\omega,p)$
does not depend on $\omega$:
$\rho_{\rm d}(\omega,p)=p^{-1}$.

Then we consider the interacting fermions. 
To analyze $\rho_{\rm d}(\omega,p)$
for interacting fermions,
the inclusion of
the nonlinearity of the fermionic
dispersion relation
is crucial~\cite{pus06},
which is not taken into account
in the Tomonaga--Luttinger 
(TL) model~\cite{tom50,lut63}.
The bosonization scheme
taking the nonlinearity into account has been
developed~\cite{pus06,teb07,ima12}
and predicted that the qualitative behavior of
$\rho_{\rm d}(\omega,p)$
drastically deviates from that in the case of free particles:
First, $\rho_{\rm d}(\omega,p)$ has 
power-law singularities at the edge of its 
support of the lower-energy side 
$\omega=\omega_{-}(p)$.
These singularities emerge due to the same mechanism 
as the singularity appearing in the X-ray
absorption rate of metals~\cite{noz69,mah00},
which is caused by the proliferation
of low-energy particle-hole pairs.
Second, $\rho_{\rm d}(\omega,p)$ exhibits
power-law suppression at 
$\omega=\omega_{+}(p)$
and the support is broaden to $\omega >\omega_{+}(p)$
because of the contribution from the multi-pair productions.
The expected shape of the strength of 
$\rho_{\rm d}(\omega,p)$
is schematically illustrated in Fig.~\ref{fig:concept_rho}(c).

\begin{figure}[!t]
	\centering
	\includegraphics[width=0.8\columnwidth]{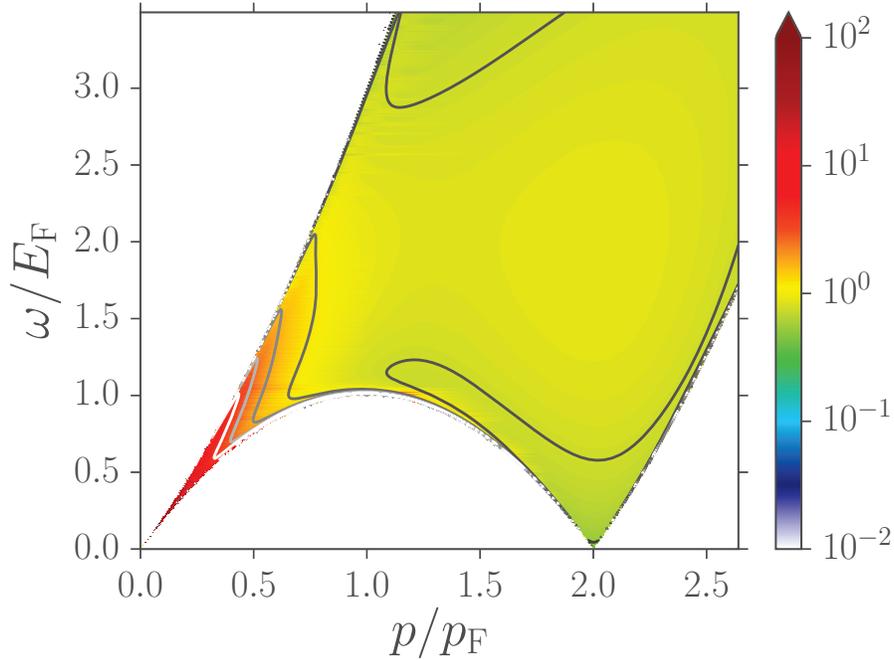}
	\caption{The contour map of the density--density
	spectral function on the $(\omega, p)$-plane.
	\label{fig:2dspectral}}
\end{figure}

Let us discuss our numerical results
of $\rho_{\rm d}(\omega,p)$.
We set the density to that at the saturation point
derived from FRG-DFT: $n=\rho_{\rm s}=1.20$ \cite{yok18}. 
Figure~\ref{fig:2dspectral} shows the contour
map of $\rho_{\rm d}(\omega,p)$
on the $(\omega, p)$-plane.
Our spectral function has the same support
as that for the free fermions,
which is in contradiction
to the expectation that
$\rho_{\rm d}(\omega,p)$ has support 
in $\omega>\omega_{+}(p)$. 
This is possibly due to the approximation made in Eq.~\eqref{eq:G4approx}, where 
the contribution from the multi-pair diagrams is discarded. 
To include the contribution from the multi-pair diagrams,
the flow of the four point correlation function
is needed to be considered, though which is beyond the scope
of this Letter.
\begin{figure}[!t]
	\centering
	\includegraphics[width=0.8\columnwidth]{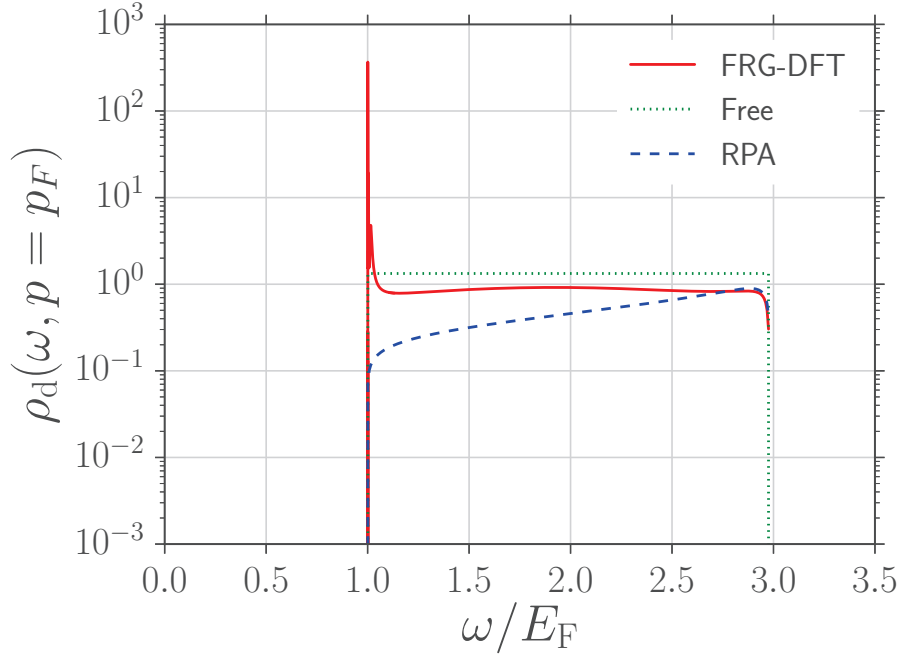}
	\caption
	{The $\omega$ dependence of the density--density 
	spectral function when the momentum is fixed to $p=p_{\rm F}$.
	The results of the FRG-DFT (solid red line),
	free particles (green dotted line) 
	and the RPA (blue dashed line) are shown.
	\label{fig:spectral}}
\end{figure}


Shown in Fig.~\ref{fig:spectral} is the spectral function 
$\rho_{\rm d}(\omega,p)$
at a fixed momentum $p=p_{\rm F}$. 
The results of the RPA and the case of free fermions are also shown 
for comparison.
The spectral function obtained by FRG-DFT 
reveals the existence of a peak 
at $\omega=\omega_{-}(p_{\rm F})$
in contrast to that derived from the RPA.
A key ingredient for such a peak to emerge
is the contribution from the second term in the right-hand 
side in Eq.~\eqref{eq:realg2flow}, 
which is not included in the RPA.
This term has singularities at
$\omega=\omega_{-}(p_{\rm F})$,
which gives the peak structure in 
$\rho_{\rm d}(\omega,p)$.
In the very close region to 
$\omega=\omega_{-}(p_{\rm})$,
we found that the spectral 
function is split into some peaks 
with slightly different energies,
which is different from 
a simple power-law singularity.

Summarizing the paper, we demonstrated how
the FRG-DFT analysis
of the ground and excited states works
in a one-dimensional continuum spinless nuclear matter.
We obtained the saturation energy
from the resultant equation of state,
which differs from that obtained using 
the Monte Carlo simulation by only 2.7\%.
Moreover, we reproduced a notable feature of
the one-dimensional fermion system that
the density--density spectral function
has singularities at the edge of its support
of the lower-energy side.
Therefore, our result suggests that the FRG-DFT is
a promising way for the analysis of not only
ground states but also excited states
of the quantum many-body systems. 
The FRG-DFT is expected to be adapted to
various systems because 
our formalism can be naturally extended to
higher-dimensional systems, and 
systems with internal degrees of freedom \cite{kem17b,ram17}, 
superfluidity, and finite temperature.

There showed up, however, 
some unexpected behaviors
in our result of 
the density--density spectral function.
At higher-energy side, the broadening
of the support of the spectral
due to the multi-pair production
has not appeared in the present framework.
This would be because
we miss the contribution from 
the multi-pair production by 
ignoring the flow of the four-point correlation function.
In addition, the spectral function was unexpectedly split into
some peaks with slightly different energies
in the very close region to
the edge of its support of the lower-energy side.
The inclusion of the flows of higher-order correlation
functions or the use of other approximation schemes 
such as the KS-FRG~\cite{lia18} is an important future work 
to see whether these are due to the approximations and/or truncations employed. 

Describing the non-uniform systems is
another interesting direction.
The introduction of a non-uniform chemical potential can realize such systems.
Our flow equations \eqref{eq:fflo}-\eqref{eq:g2fl} are straightforwardly extended
to the case of non-uniform chemical potential and can be used to
analyze non-uniform systems.

We thank Jean-Paul Blaizot for his interest in and
critical and valuable comments on the present work.
We also acknowledge Christof Wetterich, Jan M. Pawlowski and Jochen Wambach
for their interest in and fruitful comments on the present work.
T.~Y. was supported by the Grants-in-Aid for JSPS fellows
(Grant No. 16J08574).
K.~Y. was supported by the JSPS KAKENHI (Grant No. 16K17687). 
T.~K. was supported by the JSPS KAKENHI Grants (Nos. 16K05350 and 15H03663)
and by the Yukawa International Program for Quark-Hadron Sciences (YIPQS).


\end{document}